\documentclass[prd,twocolumn]{revtex4}

  \usepackage[usenames,dvipsnames]{pstricks}
\usepackage{amsmath}
\usepackage{amssymb}
\usepackage{epsfig}
\usepackage{pspicture}
\usepackage{graphicx}
\usepackage{graphics}
\usepackage{booktabs}
 \usepackage{pst-grad} % For gradients
 \usepackage{pst-plot} % For axes
 \usepackage{subfloat}

\begin{document}

\title{On Free-Electron Laser Growing Modes and their Bandwidth}

\author{Stephen Webb}
\email{swebb@grad.physics.sunysb.edu}
\affiliation{Department of Physics \& Astronomy, Stony Brook University}
\altaffiliation{Collider-Accelerator Department, Brookhaven National Laboratory}

\author{Gang Wang}
\email{gawang@bnl.gov}
\affiliation{Collider-Accelerator Department, Brookhaven National Laboratory}

\author{Vladimir Litvinenko}
\email{vl@bnl.gov}
\affiliation{Collider-Accelerator Department, Brookhaven National Laboratory}

\date{\today}

\begin{abstract}
Free-electron lasers play an increasing role in science, from generating unique femtosecond X-ray pulses for single short recording of the protein structures to amplifying feeble interactions in advanced cooling systems for high-energy hadron colliders. While modern Free-electron laser codes can describe their amplification mechanism, a deep analytical understanding of the mechanism is of extreme importance for a number of applications. Mode competition, their growth rates and amplification bandwidth are among the most important parameters of a free-electron laser. A dispersion relation, which defines these important characteristics, can be solved analytically only for a very few simple cases. In this letter we show that for a typical bell-shape energy distribution in electron beam there is no more that one growing mode. We also derive an analytical expression which determines the bandwidth of the free-electron laser.
\end{abstract}

\maketitle

Since the invention of free-electron lasers (FELs) by Madey (\cite{madey1},\cite{madey2}) the FEL field has made tremendous progress both experimentally (\cite{lcls}, \cite{efel}, \cite{jfel}) and theoretically. In an FEL electrons propagate through a wiggler, a device with a periodically oscillating transverse magnetic field, which causes the electrons to radiate at a resonant wavelength of
\begin{equation}
\lambda_r = \frac{\lambda_w}{2 \gamma_0^2} \left ( 1 + a_w^2 \right )
\end{equation}
where $a_w = e A_w/m c$ is the wiggler normalized vector potential, $\lambda_w$ is the undulator period and $\gamma = \mathcal{E}/mc^2$ is the electron relativistic factor. A resonant interaction between the electron beam and the TEM field of the radiation can become unstable, resulting in the exponential growth of the radiation power. This results in the generation of coherent radiation. Such high-gain FELs are of great interests for current and future applications, and are the subject of the discussions in this letter.

Since exact analytical solutions in three dimensions are typically intractable, numerical code such as GENESIS \cite{reiche1} and GINGER \cite{fawley1} are used for practical FEL designs. However, there do exist some analytical treatments such as \cite{ckx}, \cite{ssynima} and \cite{bpn} that can provide insight into the delicate phase-space dynamics, statistical characteristics of FEL radiation, and can be used for testing the validity of the assumptions made in the FEL codes. The FEL linear regime describing the initial amplification of the density perturbation is frequently used for an in-depth analysis of the FEL physics, and most treatments assume a wide electron beam co-propagating with the TEM wave.

Still a number of fundamental questions about FELs remain unanswered. It is understood that the self-consistent 1D FEL equation may have a variable number of modes depending upon the details of the energy distribution, with asymptotic formulae for the largest growing modes discussed in \cite{km} and \cite{ssynima} and a direct discussion of the diverging number of modes given in \cite{wl}. For example, it is known that for a beam with a gaussian energy distribution an infinite number of modes exist. But there is no answer to how many of these modes are growing, or what frequency cutoffs might exist for these growing modes.

Saldin, Schneidmiller and Yurkov provided the most general treatment of the high gain  free-electron laser operating in the small-signal, linear regime \cite{ssynima}. Using a Laplace transformation they reduced the self-consistent Maxwell-Vlasov equations \cite{vlas67} to a dispersion relation of the form
\begin{equation}
s = \frac{\hat{D}(s)}{1 - \hat{\Lambda}_p^2 \hat{D}(s)}
\end{equation}
where $s$ is the Laplace transformation variable along the longitudinal direction, $\hat{\Lambda}_p^2$ is a longitudinal space charge parameter, and $\hat{D}(s)$ is the dispersion integral given by
\begin{equation}
\hat{D}(s) = \int d\hat{P} ~ \frac{d \hat{F}}{d \hat{P}} \frac{1}{s + \imath (\hat{C} + \hat{P})}
\end{equation}
where we utilize the conventions in \cite{ssy} of $\hat{P} = (\mathcal{E} - \mathcal{E}_0)/(\rho \mathcal{E}_0)$ as the normalized energy deviation, $\hat{C} = (k_w - \omega (1 + a_w^2)/(2 c \gamma^2) ) L_G$ the normalized detuning from the FEL resonance wave number, $\hat{F}$ the normalized energy distribution of the electron bunch, the normalized to the $e$-fold gain length, $L_G$, and Pierce parameter, $\rho = (k_w L_G)^{-1}$, and where $k_w = 2 \pi / \lambda_w$. The authors also found analytical solutions for the case of the mono-energetic and Lorentzian energy distributions as roots of a cubic equation and expressed the solution for the system in the form
\begin{equation}
V(z) = \sum_{n = 1}^3 V_n ~ e^{s_n z}
\end{equation}
where the $V_n$ contain the initial conditions and the $s_n$ are the solutions of the dispersion relation. Most generally, for some energy distribution the solution is a linear superposition of exponential modes of the form $e^{s_\imath z}$ where $\{s_\imath\}$ are the solutions to the corresponding dispersion relation.

Hence, the solutions with $\Re(s_\imath) > 0$ is growing exponentially and knowing the number of these modes and their growth rate is the key for any high-gain FEL with gain much larger than unity. In this letter we will prove that for an electron beam with a typical bell-shaped energy distribution, there is no more than one growing mode, and therefore in a high-gain FEL one can use the approximation that $V(z) = V_0 e^{s_0 z}$. From the analytical results for monoenergetic and Lorentzian beams, it is already known that there is only one growing mode.

For most short wavelength FELs the longitudinal space charge does not play an important role and we neglect it in this letter, we therefore focus on the reduced dispersion relation
\begin{equation}
w(s) = s - \hat{D}(s) = 0
\end{equation}

To study this, we use the Argument Principle of complex analysis. This is an analogy of the Nyquist stability criterion (\cite{nyquist}, \cite{maccoll}); the change in the argument of a complex function $w(z)$ meromorphic on a closed contour $C$, is given by
\begin{equation}
W = Z - P = \frac{1}{2 \pi} \oint_C d \left ( \textrm{arg}(w(z)) \right )
\end{equation}
where $Z$ is the number of zeros of $w(z)$ within the contour and $P$ is the number of poles, counting multiplicity.

\begin{figure} % Change this value to rescale the drawing.
\begin{pspicture}(0,-3)(6,3)
\psline[linewidth=0.03cm]{<-}(3,-3)(3,0)
\psline[linewidth=0.03cm]{->}(3,0)(3,3)
\psline[linewidth=0.03cm]{<-}(0.0,0)(3,0)
\psline[linewidth=0.03cm]{->}(3,0)(6,0)
\psarc[linewidth=0.03, linestyle = dashed, arrowsize=0.12cm]{->}(3.1,0){2.8}{270.0}{90.0}
\psline[linewidth=0.03cm, linestyle = dashed, arrowsize=0.12cm, dash=.32cm .16cm]{<-}(3.1,-2.8)(3.1,2.8)
\usefont{T1}{ptm}{m}{it}
\rput(2.6,2.8){\small{}$\Im(s)$}
\usefont{T1}{ptm}{m}{it}
\rput(5.5,-.4){\small{}$\Re(s)$}
\psline[linewidth=0.03cm,arrowsize=0.12cm]{->}(3,0)(5, 2.1)
\usefont{T1}{ptm}{m}{it}
\rput(4, 1.3){$\textbf{R}$}
\usefont{T1}{ptm}{m}{it}
\rput(5.5,0.5){$\textbf{C}$}
\end{pspicture} 
\caption{The contour over the complex-valued $s$ that encompasses the entire right half-plane. This contour determines the number of growing roots.}
\label{windfig}
\end{figure}
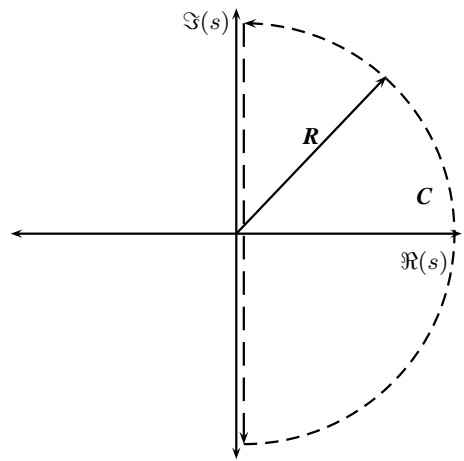

Consider the example of the contour given in figure (\ref{windfig}), which encompasses the entire right half-plane in the $R \rightarrow \infty$ limit. By calculating the winding number, $W$, of the dispersion relation along this contour, we can make a general statement about the number of growing modes.

First, we must determine the number of poles in the right half-plane of $s$. To determine this, look at the evaluation of the dispersion integral $\hat{D}(s)$, since the poles of the full dispersion relation depend upon the poles of $\hat{D}(s)$. When evaluating the Laplace transform of the linear equation for the FEL perturbation, it is assumed that $\Re(s) > 0$ by consideration of causality. Specifically,
\begin{equation}
\mathcal{L}[f(t)] = \int_0^\infty dt ~ e^{-s t} f(t)
\end{equation}
requires that, for $t>0$, $\Re(s) > 0$ for the integral to converge. When applied to the evaluation of $\hat{D}(s)$ by integrating over the complex $\hat{P}$-plane, this means that the contour is given by figure (\ref{dispinta}). The correct analytic continuation is in figure (\ref{dispintb}) as first shown in \cite{llevib}.

% Generated with LaTeXDraw 2.0.8
% Sat Apr 09 22:25:29 EDT 2011
% \usepackage[usenames,dvipsnames]{pstricks}
% \usepackage{epsfig}
% \usepackage{pst-grad} % For gradients
% \usepackage{pst-plot} % For axes
\begin{subfigures}
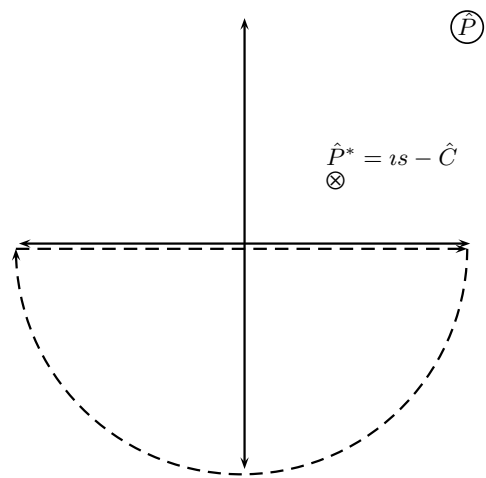
\begin{figure}
\scalebox{1} % Change this value to rescale the drawing.
{
\begin{pspicture}(0,-2)(4,4)
\rput(2.04,1.07){\psaxes[linewidth=0.03,arrowsize=0.05291667cm 2.0,arrowlength=1.4,arrowinset=0.4,labels=none,ticks=none,ticksize=0.10583333cm]{<->}(0,0)(-3,-3)(3,3)}
\psline[linewidth=0.03cm,linestyle=dashed,dotsep=0.16cm,arrowsize=0.05291667cm 2.0,arrowlength=1.4,arrowinset=0.4]{->}(-1,1)(5,1)
\psarc[linewidth=0.03,linestyle=dashed,dotsep=0.16cm,arrowsize=0.05291667cm 2.0,arrowlength=1.4,arrowinset=0.4]{<-}(2,1){3}{180}{0}
\psdots[dotsize=0.24,dotstyle=otimes](3.25,1.91)
\psdots[dotsize=0.45,dotstyle=o](5,3.95)
\rput(5,4){$\hat{P}$}
\rput(4,2.25){$\hat{P}^* = \imath s - \hat{C}$}
\end{pspicture} 
}
\caption{Contour integration for $\hat{D}(s)$ with $\Re(s) > 0$.}
\label{dispinta}
\end{figure}
\begin{figure}
\scalebox{1} % Change this value to rescale the drawing.
{
\begin{pspicture}(0,-2)(4,4)
\rput(2.04,1.07){\psaxes[linewidth=0.03,arrowsize=0.05291667cm 2.0,arrowlength=1.4,arrowinset=0.4,labels=none,ticks=none,ticksize=0.10583333cm]{<->}(0,0)(-3,-3)(3,3)}
\psline[linewidth=0.03cm,linestyle=dashed,dotsep=0.16cm,arrowsize=0.05291667cm 2.0,arrowlength=1.4,arrowinset=0.4]{->}(-1,1)(3,1)
\psline[linewidth=0.03cm,linestyle=dashed,dotsep=0.16cm,arrowsize=0.05291667cm 2.0,arrowlength=1.4,arrowinset=0.4]{->}(3.5,1)(5,1)
\psline[linewidth=0.03cm,linestyle=dashed,dotsep=0.16cm,arrowsize=0.05291667cm 2.0,arrowlength=1.4,arrowinset=0.4]{<-}(3.5,1)(3.5,0)
\psline[linewidth=0.03cm,linestyle=dashed,dotsep=0.16cm,arrowsize=0.05291667cm 2.0,arrowlength=1.4,arrowinset=0.4]{->}(3,1)(3,0)
\psarc[linewidth=0.03,linestyle=dashed,dotsep=0.16cm,arrowsize=0.05291667cm 2.0,arrowlength=1.4,arrowinset=0.4]{<-}(2,1){3}{180}{0}
\psarc[linewidth=0.03,linestyle=dashed,dotsep=0.16cm,arrowsize=0.05291667cm 2.0,arrowlength=1.4,arrowinset=0.4]{->}(3.25,0){.25}{180}{0}
\psdots[dotsize=0.24,dotstyle=otimes](3.25,0)
\psdots[dotsize=0.45,dotstyle=o](5,3.95)
\rput(5,4){$\hat{P}$}
\rput(3.5,-.5){$\hat{P}^* = \imath s - \hat{C}$}
\end{pspicture} 
}
\caption{Contour integration for $\hat{D}(s)$ for the proper analytic continuation of $\Re(s)$ over the entire complex $s$-plane.}
\label{dispintb}
\end{figure}
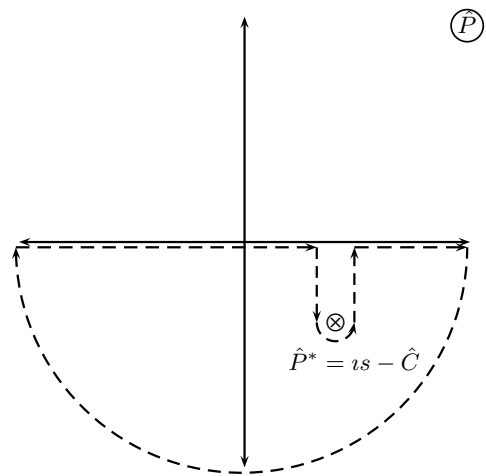
\end{subfigures}

The pole at $\imath s - \hat{C}$ must remain above the contour integration when the dispersion relation is analytically continued to deal with the case of $\Re(s) \leq 0$. This causality requirement means that the contour remains in the lower half-plane of $\hat{P}$, and therefore the poles of $\hat{D}(s)$ are located at points given by the form
\begin{equation}
s = -\imath (\hat{C} + \hat{P}^*_n)
\end{equation}
for a $\hat{P}^*_n$ with an imaginary part less than zero. This translates to the poles of $\hat{D}(s)$ all existing in the left half-plane, so there are no poles inside our contour and $P=0$. Therefore the number of growing modes is given by the winding number of the dispersion relation! Causality in the Laplace transform is therefore directly related to the pole structure of the dispersion relation, which in turn gives us useful information about the number of amplifying modes.

From very general considerations of the shape of $\hat{F}$, we can arrive at a very solid schematic understanding of how the contour maps. The arc at infinity we parameterize by $s = R e^{\imath \theta}$ for $\theta \in (-\pi/2, \pi/2]$. Along this portion of the contour, $\hat{D} \sim R^{-2}$ and the dispersion relation is the identity map.

Along the vertical end of the contour, $s = \imath t$ for $t \in (\infty, - \infty)$ and the dispersion integral is given by
\begin{equation}\label{dispeqn}
\begin{split}
 \int d\hat{P} \frac{d \hat{F}}{d \hat{P}} \frac{1}{ \imath (t + \hat{C} + \hat{P})} = \\ \mathcal{P} \int d\hat{P} \frac{d \hat{F}}{d \hat{P}} \frac{1}{ \imath (t + \hat{C} + \hat{P})} + \pi \hat{F}' (\hat{P} = - t - \hat{C}) 
\end{split}
\end{equation}
where we have used the well-known identity
\begin{equation}
[\cdots]\frac{1}{x} dx = \mathcal{P} \left ([\cdots] \frac{1}{x} \right ) dx + \imath \pi [\cdots] \delta(x) dx
\end{equation}
where $\mathcal{P}$ indicates the Cauchy Principal Value. The first term is purely imaginary since $\hat{F}$ and $(t + \hat{C} + \hat{P})^{-1}$ are real functions, and the second term is pure real. The real part of the above expression is very simple, and provides information when the mapping crosses the imaginary axis, \emph{i.e.} this happens when $\hat{F}' = 0$. As a first example, we consider a simple bell curve shape for $\hat{F}$, a curve with a single extremum $\hat{F}' = 0$ at $t + \hat{C} = t_0$.

In this case the map of the vertical line crosses the imaginary axis, $\Re(\imath t - \hat{D}(\imath t) )  = 0$ once and only once. Furthermore, because $\hat{F}$ has a bell shape, at the local maximum the second derivative of $\hat{F}$ -- the derivative of the real part of the contour -- is negative. Therefore, the contour is crossing from the right half-plane to the left half-plane. This leaves two possible options for the contour, given in figure (\ref{conta}-\ref{contb}). The contour which crosses above the real axis corresponds to a winding number of one, while the contour which crosses below the real axis corresponds to a winding number of zero. Hence the contour in figure \ref{conta} corresponds to a single growing mode, and the contour in figure \ref{contb} corresponds to the case when all modes are either oscillating or decaying. Thus, we prove that for a typical bell-shaped energy distribution the one-dimensional FEL has one or no growing modes.

\begin{subfigures}
% Generated with LaTeXDraw 2.0.8
% Sun Apr 10 18:14:40 EDT 2011
% \usepackage[usenames,dvipsnames]{pstricks}
% \usepackage{epsfig}
% \usepackage{pst-grad} % For gradients
% \usepackage{pst-plot} % For axes
\begin{figure}
\scalebox{1} % Change this value to rescale the drawing.
{
\begin{pspicture}(0,-3.2)(6.1,3.2)
\rput(3.02,0.04970601){\psaxes[linewidth=0.04,arrowsize=0.05291667cm 2.0,arrowlength=1.4,arrowinset=0.4,labels=none,ticks=none,ticksize=0.10583333cm]{<->}(0,0)(-3,-3)(3,3)}
\psarc[linewidth=0.04,linestyle=dashed,dash=0.16cm 0.16cm](3.04,0.08970601){3.04}{270.0}{90.93919}
\psbezier[linewidth=0.04,linestyle=dashed,dash=0.32cm 0.16cm](3.0,3.069706)(3.0,2.269706)(4.157667,2.9537966)(3.98,1.969706)(3.8023329,0.9856154)(1.8151292,1.5918782)(1.48,0.649706)(1.1448708,-0.29246616)(2.0200844,-3.129706)(3.0,-2.930294)
\rput(5,3){$w(s)$}
\end{pspicture} 
}
\caption{Contour for a growing mode for a bell curve distribution.}
\label{conta}
\end{figure}
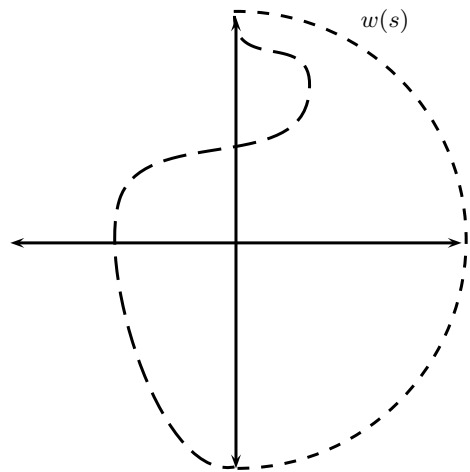
\begin{figure}
% Generated with LaTeXDraw 2.0.8
% Sun Apr 10 18:17:30 EDT 2011
% \usepackage[usenames,dvipsnames]{pstricks}
% \usepackage{epsfig}
% \usepackage{pst-grad} % For gradients
% \usepackage{pst-plot} % For axes
\scalebox{1} % Change this value to rescale the drawing.
{
\begin{pspicture}(0,-3.2)(6.1,3.2)
\rput(3.02,0.08419542){\psaxes[linewidth=0.04,arrowsize=0.05291667cm 2.0,arrowlength=1.4,arrowinset=0.4,labels=none,ticks=none,ticksize=0.10583333cm]{<->}(0,0)(-3,-3)(3,3)}
\psarc[linewidth=0.04,linestyle=dashed,dash=0.16cm 0.16cm](3.04,0.12419542){3.04}{270.0}{90.93919}
\psbezier[linewidth=0.04,linestyle=dashed,dash=0.32cm 0.16cm](3.02,3.0841954)(3.02,2.2841954)(4.9153104,0.8510946)(4.54,-0.075804584)(4.1646895,-1.0027038)(2.2092495,-0.11156159)(1.88,-1.0558046)(1.5507506,-2.0000477)(2.03669,-3.1641953)(3.0,-2.8958046)
\rput(5,3){$w(s)$}
\end{pspicture} 
}
\caption{Contour for zero growing modes.}
\label{contb}
\end{figure}
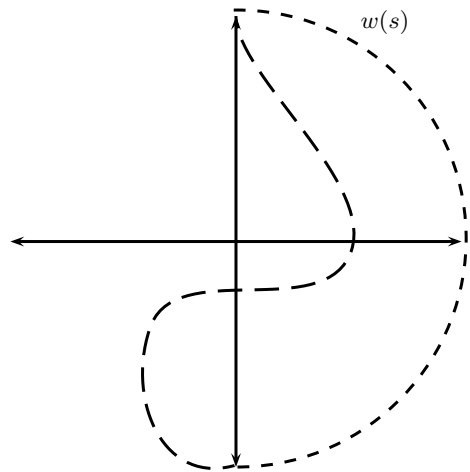
\end{subfigures}

Because the contour crosses the imaginary axis at $t  + \hat{C} = t_0$, the location along the imaginary axis, either above or below zero, depends upon the detuning. Thus, by inserting this solution into the imaginary part of the dispersion function, equation (\ref{dispeqn}), we can write an expression for the detuning at which the growth rate of the FEL vanishes:
\begin{equation}\label{bandwidth}
\hat{C}^* = \Im \left [\imath t_0 +  \int d\hat{P} ~ \frac{d \hat{F}}{d \hat{P}} \frac{1}{\imath (t_0 + \hat{P})} \right ]
\end{equation}
For a gaussian distribution given by
\begin{equation}
\hat{F}(\hat{P}) = \frac{1}{\sqrt{2 \pi \sigma_P^2}} \exp \left \{ - \hat{P}^2/(2 \sigma_P^2) \right \}
\end{equation}
this gives the critical detuning as
\begin{equation}
\hat{C}^* = - \frac{1}{\sigma_P^2}
\end{equation}
More generally, it is clear by dimensional analysis that for any energy distribution with only one scale $\sigma_P$ the critical frequency scales as $\hat{C}^* \propto \sigma_P^{-2}$ with the numerical quantity on the order of unity and dependent on the particulars of $\hat{F}'(\hat{P})$.

To summarize this result, assuming only that $\hat{F}(\hat{P})$ has a single maximum and goes to zero at $\hat{P} \rightarrow \pm \infty$, we have shown that causality demands a single mode which amplifies frequencies with $\hat{C} > \hat{C}^* \propto - \sigma_P^{-2}$ with some numerical quantity depending on the specifics of the distribution and $\sigma_P$ is the energy spread parameter for the distribution. We should note that this result holds even for the cold beam with zero energy spread, which amplifies all frequencies above the FEL resonance frequency (note that negative $\hat{C}$ corresponds to higher frequencies).

It is worth mentioning that, generally speaking, an FEL with an energy distribution with $N$ humps can have up to $N$ growing modes. It also means that for any typical energy distribution there is a finite number of growing modes. The arguments used to construct the contours in figures (\ref{conta}-\ref{contb}) may also be used to draw schematics for a multiple-peak energy distribution. Every local maximum represents the contour crossing from the right half-plane to the left half-plane and every local minimum represents the contour crossing from the left to the right half-plane. Extrema that are neither maxima nor minima (i.e. plateaus) are bounces and do not contribute. By looking at the sign flips in the imaginary part, analogous diagrams can be drawn and by a similar argument there are zero, one, or two growing modes corresponding to one mode for each resonant frequency that corresponds to a local maximum, and having one, both, or none of them being amplified at a given frequency. Similar criterion to the above for a critical detuning may be developed using the same arguments. In practice this is of little real interest, but it is intellectually satisfying that this formalism can consider multiple resonant frequencies caused by multiple peaks in the energy distribution. Because the specifics of the number of growing modes is independent of the energy distribution beyond some basic considerations on the shape of the distribution, we argue that this result may be regarded as topological.

The method used to obtain this result does not depend on the fact that we are dealing with an FEL, and indeed these results have been obtained in frequency space by Nyquist \cite{nyquist}, Penrose for a stationary plasma driven by an external electric field with frequency $\omega$ \cite{penrose}, and for general ``growing waves" \cite{sturrock} or other plasma instabilities \cite{briggs}. This particular method appears to be periodically rediscovered by physicists over the decades. Because the Laplace transform and Fourier transform are related by a Wick rotation, the same formal structure exists either looking at the initial value problem or at frequency responses.

Therefore we have proven that, for a reasonable bell-shaped type energy distribution in the electron beam, the FEL dispersion relation has only one growing solution, which exists only within a frequency range defined by a simple expression (\ref{bandwidth}). We extend our qualitative topological considerations for energy distributions having multiple peaks and conclude that for a smooth and finite energy distribution there can only be a finite number of growing modes, simply related to the number of extrema of the energy distribution.

The authors would like to thank Michael Blaskiewicz and Geoffrey Krafft for helpful discussions. Work supported by Brookhaven Science Associates, LLC under Contract No. DE-AC02-98CH10886 with the U.S. Department of Energy.

\bibliography{dispersionbib}

\end{document}